\documentclass[12pt, a4paper]{article}

\usepackage[utf8]{inputenc}
\usepackage{amsmath}
\usepackage{amsfonts}
\usepackage{amssymb}
\usepackage{graphicx}
\usepackage[margin=1in]{geometry}
\usepackage{booktabs} 
\usepackage{hyperref} 
\usepackage{natbib}   
\usepackage{filecontents} 

\hypersetup{
    colorlinks=true,
    linkcolor=blue,
    filecolor=magenta,      
    urlcolor=cyan,
    citecolor=blue,
}

\title{Mitigating Financial Frictions in Agriculture: A Framework for Stablecoin Adoption}
\author{Xinyu Li}
\date{\today}

\begin{document}

\maketitle

\begin{abstract}
\noindent Persistent financial frictions—including price volatility, constrained credit access, and supply chain inefficiencies—have long hindered productivity and welfare in the global agricultural sector. This paper provides a theoretical and applied analysis of how fiat-collateralized stablecoins, a class of digital currency pegged to a stable asset like the U.S. dollar, can address these long-standing challenges. We develop a farm-level profit maximization model incorporating transaction costs and credit constraints to demonstrate how stablecoins can enhance economic outcomes by (1) reducing the costs and risks of cross-border trade, (2) improving the efficiency and transparency of supply chain finance through smart contracts, and (3) expanding access to credit for smallholder farmers. We analyze key use cases, including parametric insurance and trade finance, while also considering the significant hurdles to adoption, such as regulatory uncertainty and the digital divide. The paper concludes that while not a panacea, stablecoins represent a significant financial technology with the potential to catalyze a paradigm shift in agricultural economics, warranting further empirical investigation and policy support. \\

\noindent \textbf{Keywords:} Stablecoins, Agricultural Economics, FinTech, Supply chain finance, Financial inclusion, Risk management. \\

\end{abstract}

\clearpage
\tableofcontents
\clearpage

\section{Introduction}

The agricultural sector, particularly in developing economies, is characterized by a persistent productivity gap \citep{gollin2014agricultural} and systemic financial exclusion. Farmers, especially smallholders, face a confluence of challenges: high price and yield volatility, limited access to affordable credit, and opaque, inefficient supply chains. These financial frictions not only suppress farm-level income but also impede capital investment, technology adoption, and overall economic development. Traditional financial instruments and institutions have often failed to adequately address these issues due to high transaction costs, information asymmetries, and significant counterparty risk.

The advent of financial technology (FinTech) offers a new frontier of potential solutions. Among the most promising innovations are stablecoins—digital assets designed to maintain a stable value relative to a target peg, typically a major fiat currency like the U.S. Dollar. Unlike volatile cryptocurrencies such as Bitcoin, stablecoins combine the technological benefits of blockchains (e.g., programmability, transparency, low-cost transfers) with the stability of traditional money \citep{narayanan2016bitcoin}. This unique combination positions them as a powerful tool for upgrading the financial infrastructure of the agricultural economy.

This paper seeks to answer the following research question: \textit{What is the theoretical potential for stablecoins to mitigate key financial frictions in agricultural markets, and what are the practical pathways and challenges to their adoption?} To address this, we first develop a theoretical framework based on a farm-level profit maximization model to formally analyze how stablecoin adoption can influence economic decision-making by reducing transaction costs and easing credit constraints. Second, we explore concrete applications, including cross-border trade, automated supply chain finance, and parametric insurance. Finally, we provide a balanced discussion of the significant challenges, including regulatory hurdles and the digital divide, that must be overcome for this potential to be realized.

Our contribution is twofold. We provide one of the first formal economic analyses connecting stablecoin architecture to agricultural finance, moving beyond descriptive accounts. Furthermore, by outlining specific mechanisms and proposing an empirical agenda, we lay the groundwork for future research to test these theoretical propositions.

\section{Literature Review}

\subsection{Financial Frictions in Agricultural Economics}
The literature on agricultural finance extensively documents the barriers faced by farmers. Access to credit is a central theme, with seminal work by \citet{stiglitz1981credit} explaining how information asymmetries lead to credit rationing, a phenomenon acutely felt in rural markets where lenders lack reliable data on borrower creditworthiness. The "market for lemons" problem \citep{akerlof1970market} is also applicable, as lenders struggle to distinguish between high- and low-risk farming operations, leading to prohibitively high interest rates for all.

Risk management is another critical area. Farmers contend with both production risk (e.g., weather, pests) and market risk (e.g., price volatility). While instruments like crop insurance exist, they often suffer from high administrative costs, moral hazard, and adverse selection. Area-yield insurance models were proposed as a partial solution \citep{miranda1991area}, but basis risk remains a significant issue.

Finally, supply chains in agriculture are notoriously complex and inefficient. Transactions between farmers, aggregators, processors, and retailers involve long settlement times, multiple intermediaries, and significant paperwork, all of which increase costs and reduce the farmgate price. These transaction costs are a central focus of institutional economics \citep{williamson1985economic} and represent a direct impediment to market efficiency.

\subsection{The Architecture and Economics of Stablecoins}
Stablecoins are a rapidly evolving class of digital assets. They can be broadly categorized into three types:
\begin{enumerate}
    \item \textbf{Fiat-Collateralized:} These are the most common and straightforward. For each stablecoin token issued, the issuer holds an equivalent amount of fiat currency (e.g., USD) in a verified bank account. Examples include USDC (Circle) and USDT (Tether). Their stability hinges on the transparency and quality of their reserves.
    \item \textbf{Crypto-Collateralized:} These are backed by a basket of other cryptocurrencies. To absorb the volatility of the underlying collateral, they are typically over-collateralized. MakerDAO's DAI is a prominent example.
    \item \textbf{Algorithmic:} These attempt to maintain their peg using algorithms that automatically expand or contract the token supply in response to market price, akin to a central bank's open market operations. This category has proven to be the most fragile.
\end{enumerate}
For the purposes of agricultural applications, fiat-collateralized stablecoins are the most relevant due to their lower complexity and higher perceived reliability. Their key economic feature is the ability to function as a programmable dollar on a blockchain. This enables "smart contracts"—self-executing contracts with the terms of the agreement directly written into code. These contracts can automate complex transactions, reducing the need for costly intermediaries and enforcement mechanisms. However, the stability of even the largest stablecoins is a subject of intense academic and regulatory scrutiny \citep{griffin2020stablecoins, fsb2020regulation}.

\section{Theoretical Framework}
To formalize the economic impact of stablecoins, we consider a representative farmer's profit maximization problem. The farmer chooses levels of a variable input, $X$ (e.g., labor, fertilizer), to produce a single crop.

\subsection{The Baseline Model}
The farmer's profit, $\pi$, is given by:
\begin{equation}
\max_{X} \quad \pi = E[p \cdot f(X; \theta)] - (w + \tau_i)X - C_f
\label{eq:profit_base}
\end{equation}
where:
\begin{itemize}
    \item $p$ is the stochastic market price of the output.
    \item $f(X; \theta)$ is the production function, with $f' > 0, f'' < 0$. $\theta$ is a stochastic variable representing weather and other yield shocks.
    \item $w$ is the unit cost of the input $X$.
    \item $\tau_i$ represents the per-unit transaction costs associated with acquiring inputs (e.g., financing costs, payment fees).
    \item $C_f$ represents fixed transaction costs associated with selling the final product (e.g., currency conversion fees, settlement delays, counterparty risk).
\end{itemize}
The first-order condition for an interior solution is:
\begin{equation}
E[p \cdot f'(X; \theta)] = w + \tau_i
\label{eq:foc_base}
\end{equation}
The farmer equates the expected marginal revenue product of the input to its effective marginal cost.

\subsection{Introducing Stablecoins}
The adoption of a stablecoin-based financial system can affect the model's parameters in several ways. Let variables with a superscript $S$ denote the state with stablecoins.

\paragraph{Reduction in Transaction Costs.} Stablecoins can dramatically reduce transaction costs for both inputs and outputs. Payments for inputs can be made instantly and with near-zero fees, reducing $\tau_i$. For sales, especially cross-border, stablecoins eliminate costly currency conversion fees and the need for correspondent banks. This reduces the fixed cost component, $C_f$.

The new profit function becomes:
\begin{equation}
\pi^S = E[p \cdot f(X; \theta)] - (w + \tau_i^S)X - C_f^S
\label{eq:profit_stable}
\end{equation}
where $\tau_i^S < \tau_i$ and $C_f^S < C_f$. The new first-order condition is:
\begin{equation}
E[p \cdot f'(X^S; \theta)] = w + \tau_i^S
\label{eq:foc_stable}
\end{equation}
Since the right-hand side is lower, and assuming diminishing marginal returns ($f'' < 0$), the optimal input use $X^S$ will be higher than the baseline $X^*$.
\\[1em]
\textbf{Proposition 1:} \textit{The adoption of stablecoins, by reducing input and output transaction costs, leads to higher optimal input use and increased farm-level profit, ceteris paribus.}

\paragraph{Easing Credit Constraints.} Smallholders are often credit constrained, meaning they cannot borrow enough to reach the optimal input level $X^*$. Let $K_{max}$ be the maximum capital available for inputs. The constraint is $(w + \tau_i)X \le K_{max}$. If this constraint is binding, the farmer uses $X_{con} = K_{max} / (w + \tau_i)$.

Stablecoins can ease this constraint through two channels:
\begin{enumerate}
    \item \textbf{Lower Cost of Capital:} By enabling peer-to-peer lending platforms and providing lenders with a transparent, real-time record of a farmer's transactions (cash flow), stablecoins can reduce the information asymmetry that leads to high risk premia. This lowers the interest rate $r$ embedded in $\tau_i$.
    \item \textbf{Increased Capital Availability:} Supply chain finance models (discussed below) can unlock working capital by allowing farmers to borrow against confirmed orders or delivered goods, increasing $K_{max}$.
\end{enumerate}

With stablecoins, the new constraint becomes $(w + \tau_i^S)X \le K_{max}^S$. Since $\tau_i^S < \tau_i$ and potentially $K_{max}^S > K_{max}$, the new constrained input level $X_{con}^S$ is unambiguously higher.
\\[1em]
\textbf{Proposition 2:} \textit{For credit-constrained farmers, stablecoins can increase productivity by lowering the effective cost of inputs and increasing the total amount of available capital.}

\section{Applications and Use Cases}

\subsection{Cross-Border Trade Finance}
Consider a coffee cooperative in Colombia exporting to a buyer in Germany. The traditional process involves letters of credit, multiple banks, and settlement times that can take weeks, with currency conversion fees at each step. By using a USD-pegged stablecoin (e.g., USDC), the German buyer can transfer funds directly to the cooperative's digital wallet. The transaction settles in minutes, not weeks, and the funds are immediately available to the cooperative in a stable currency, which they can then convert to local currency as needed. This reduces working capital cycles and foreign exchange risk.

\subsection{Automated Supply Chain Finance}
Smart contracts on a blockchain can automate the entire procure-to-pay process.
\begin{enumerate}
    \item A processor places a purchase order for 10 tons of wheat from a farmer, lodging the payment in a stablecoin-based smart contract.
    \item The farmer delivers the wheat to a designated silo. An IoT sensor or a trusted third-party agent (an "oracle") verifies the quantity and quality and writes this data to the blockchain.
    \item Upon verification, the smart contract automatically releases the payment to the farmer's wallet.
\end{enumerate}
This system drastically reduces counterparty risk for the farmer and administrative overhead for the processor. It also creates a verifiable track record of delivery, which can be used to secure financing against future orders.

\subsection{Parametric Insurance}
Traditional crop insurance is plagued by high assessment costs and delays. A parametric insurance product can be built using a smart contract and stablecoins.
\begin{itemize}
    \item Farmers pay premiums into a smart contract pool.
    \item The contract is linked to a trusted, publicly verifiable data source, such as satellite rainfall data from NOAA.
    \item The trigger is defined in the contract: e.g., "If cumulative rainfall in region Y is less than Z mm between April 1 and June 30, execute payout."
    \item If the trigger condition is met, the smart contract automatically distributes stablecoin payouts to all insured farmers in the affected region.
\end{itemize}
This eliminates claim assessment costs and moral hazard, and delivers liquidity to farmers immediately after a shock when it is most needed.

\section{Challenges and Policy Implications}
Despite the immense potential, the path to adoption is fraught with challenges.
\begin{itemize}
    \item \textbf{Regulatory Uncertainty:} The legal status of stablecoins is ambiguous in many jurisdictions. Policymakers must create clear frameworks that balance innovation with financial stability and consumer protection \citep{fsb2020regulation}.
    \item \textbf{The Digital Divide:} These solutions require internet access and a basic level of digital literacy, which are often lacking in the most remote agricultural communities. This risks exacerbating existing inequalities. Public investment in rural digital infrastructure and education is paramount.
    \item \textbf{Systemic and Operational Risks:} A major fiat-collateralized stablecoin could "break the buck" if its reserves are mismanaged, causing systemic disruption \citep{griffin2020stablecoins}. Furthermore, smart contracts are vulnerable to bugs and hacks, and secure private key management is a challenge for many users.
    \item \textbf{Interoperability:} The current landscape is fragmented across different blockchains and stablecoins. Lack of interoperability creates silos and friction, undermining the core benefits.
\end{itemize}
Addressing these challenges requires a concerted effort from technologists, policymakers, and development organizations. Pilot projects are needed to test feasibility and build trust. The success of mobile money platforms like M-PESA in Kenya provides a valuable precedent for how technology can be adapted to local contexts and achieve widespread adoption \citep{jack2011mobile}.

\section{Conclusion and Avenues for Future Research}
This paper has argued that stablecoins are not merely a speculative digital asset but a potentially transformative technology for agricultural economics. By providing a stable, programmable, and efficient medium of exchange, they can directly address the core financial frictions—high transaction costs, credit constraints, and risk—that have long suppressed agricultural productivity. Our theoretical model shows that reducing these frictions leads to higher input use and profits, particularly for credit-constrained smallholders.

The analysis highlights a clear and compelling potential. However, the gap between theoretical potential and on-the-ground reality is significant. Future research should move towards empirical validation. A promising avenue would be a quasi-experimental approach, such as a difference-in-differences analysis comparing agricultural cooperatives or supply chains that participate in a stablecoin-based pilot program with a matched control group. Key outcome variables to measure would include transaction costs, loan interest rates, input usage, farmgate prices, and income levels.

Furthermore, research is needed on the governance of these new financial ecosystems. Who designs the smart contracts? Who are the trusted oracles? How can smallholder farmers have a voice in the governance of platforms that are becoming central to their livelihoods? Answering these questions is critical to ensuring that the digital dollar in the dirt cultivates broad-based prosperity, not new forms of digital dependency.

\bibliographystyle{apalike}
\bibliography{references}

\end{document}